\documentstyle[epsf,psfig]{mn}     

\newcommand{\beq}{\begin{equation}}
\newcommand{\eeq}{\end{equation}}

\begin{document}
\title
[Stellar population in  AGN]
{Stellar population in Active Galactic Nuclei. I. The Observations\thanks{Based on observations collected at the 
Canadian-French-Hawaiian Telescope,
Hawaii, and Observatoire de Haute Provence, France.} }
               
   \author
    [M. Serote Roos et al.]
    {M. Serote Roos$^{1,2}$, C. Boisson$^1$, M. Joly$^1$, M. J. Ward$^3$ \\  
     $^1$DAEC, Unit\'e associ\'ee au CNRS et \`a l'Universit\'e Denis Diderot\\
     Observatoire de Paris, section de Meudon\\ 
     92195 Meudon Cedex, France\\
     $^2$Observat\'orio Astron\'omico de Lisboa\\
     Tapada da Ajuda\\
     1300 Lisboa, Portugal\\
     $^3$ X-ray Astronomy Group, Department of Physics and Astronomy\\ 
     University of Leicester\\
     Leicester LE1 7RH, England}

 \date{accepted : december 1997}
\maketitle

\begin{abstract}
   Recent observations supported by theoretical models have lead to the
view that giant and supergiant stars are over abundant, and/or a
high metallicity component may be present, in the stellar
populations at the centres of active galaxies. Here we attempt to
quantify these effects by observing the strengths of the stellar
absorption lines of Mg~b, NaI, CaII triplet as well as molecular
bands such as CN and TiO. Using long-slit spectroscopic data we are
able to separate the stellar populations in and around the nucleus,
for a sample including, normal, LINER, starburst and Seyfert galaxies.
 
In this paper we present the data, namely spectra of the nucleus and of 
a number of circum-nuclear regions. Comparisons reveal gradients in
both the reddening and the stellar population within the central
regions of most galaxies. Detailed stellar population synthesis is
presented in a companion paper.

\end{abstract}

\begin{keywords}
  Galaxies: stellar content, active nuclei
  -- methods: data analysis
  -- methods: stellar populations
\end{keywords}
\section{Introduction}

A crucial unsolved question is whether the stellar populations in the
nuclear regions of Active Galaxies differ from those of non-active
galaxies of the same Hubble type. Correlations between near IR CO
indices, far-infrared and X-ray luminosities of active galaxies may
indicate that the more powerful monsters reside in more actively
star-forming host galaxies (Yamada, 1994). The accelerated star
formation caused by dynamic instabilities which trigger and/or fuel
the nuclear activity could result in an overabundance of giant,
supergiant and super metal rich (SMR) stars (Scoville, 1992 and
references therein).
  
Terlevich et al. (1990) observed that in some Active Galactic Nuclei
(AGN) the near IR CaII triplet absorption features are as strong, or
even stronger than those of normal non-active galactic nuclei. They
have suggested that the ``featureless" blue continuum previously
thought to be non-stellar in origin, actually arises from the
unresolved continuum from a young cluster of stars containing red
supergiants. The presence of a ``truly'' featureless continuum is then
required only in the case of some Seyfert~1 galaxies which show
dilution of the CaII triplet lines. Based on independent evidence we
know that in some galaxies a starburst region surrounds the unresolved
nucleus which emits the broad lines e.g. NGC~7469 (Wilson et al.,
1991). Nonetheless this does not prove a causal connection between the
starburst and AGN.
  
An alternative interpretation, plausible within the framework of our
knowledge of AGN, is that the stellar lines are superimposed on a
non-stellar nuclear continuum, and are formed in a super metal rich
population. Indeed, one could expect abundance anomalies due to the
intense star formation in the metal rich environment of the nucleus.

 From our detailed study of the Seyfert~1 nucleus in the galaxy
NGC~3516 (Serote Roos et al., 1996), we find that the stellar
population exhibits a noticeable dilution by a featureless continuum
in the wavelength range 5000-9800\AA\ {\it as well as} a high
proportion of super metal rich stars.

Thus, before dismissing the possibility of the presence of any
non-thermal component in the near infrared continuum of Seyfert
galaxies, we must first define the stellar population in the nucleus
and the surrounding regions for a sample of AGN of all levels of
activity.

 Until now most studies tackling this subject have made use of only a
small spectral domain around a few spectral features. In order to make
further progress it is necessary to extend these studies to cover more
stellar absorption features, providing signatures of different stellar
populations e.g. MgI $\lambda$5175, NaI~D $\lambda$5896 and 8196\AA\ 
and the numerous TiO and CN bands. The NaI lines, for example,
help to distinguish between the effects of overmetallicity rather than
a supergiant dominated population. Indeed, the ambiguity between
metallicity and supergiant excess can be resolved by comparison of the
CaII triplet and NaI absorption strengths (e.g. Zhou, 1991) although the
interstellar contribution to NaI introduces some uncertainty, and also
by comparison with the strength of MgI (Couture \& Hardy, 1990).

In this paper we present long-slit spectroscopy, in the range
5000-9800\AA\, of a sample of galaxies with different levels of
activity. These observations are used to estimate radial gradients in
the stellar population and to extract the stellar spectrum of the very
nucleus.  The sample is selected to include various classes of active
galaxies, i.e. Seyferts 1 and 2, LINERs and starbursts, in order that
the strength of the stellar features may be correlated with the
general properties of each class.  We also present data for two
normal galaxies which will be used as comparison templates. This is
not a complete sample in any statistical sense, but it does provide
insights into the diversity of bulge stellar populations in and around
AGN of different classes of activity.

The detailed analysis of the stellar populations for this sample using
a new spectral synthesis programme (Pelat, 1997), is to be found in
Paper II (Serote Roos et al., 1997).

\section{Observations and data analysis}

The observations of 15 galaxies were carried out in January 1990, at
the CFHT, Hawaii. Table~1 gives a list of the galaxies, together with
their morphological type, level of activity, radial velocity, distance
in Mpc (assuming H$_o$=75km/s/Mpc and q$_o$=0.0) and colour excess due
to galactic interstellar reddening. Distances have been derived from
the observed emission line redshift except for the nearby galaxies
NGC~3310, NGC~3379, NGC~3521, NGC~4278 and M~81 for which distances are
based on published recessionnal velocities and Virgocentric infall 
(Kennicutt, 1988; Bender et al., 1992). Values of E(B-V) have been
evaluated using the galactic hydrogen column density derived from the
21cm survey of Stark et al.  (1984) and Elvis et al.  (1989).

\begin{table*}
\caption[]{Observed galaxies.}
\bigskip
\label{galaxias}
\begin{tabular}{|l|c|c|c|c|c|}
\hline
\noalign{\smallskip}
\multicolumn{1}{c}{galaxy}&morphology&\multicolumn{1}{c}{activity level}&\multicolumn{1}{c}{radial velocity} & D &\multicolumn{1}{c}{E(B-V)}\\
 & & &\multicolumn{1}{c}{(km s$^{-1}$)}&Mpc &\\
\noalign{\smallskip}
\hline
\noalign{\smallskip}

NGC 1275 & E cD   & Seyfert 1 &5264&72 &0.20\\
NGC 2110 &E3 or S0& Seyfert 2 &2284&32 &0.26\\
NGC 3227 & Sb     & Seyfert 1 &115&12 &0.03\\
NGC 3310 & Sbc    & starburst &980&19  &0.01\\
NGC 3379 & E 3    & normal    &920&8  &0.04\\
NGC 3516 & SB0a   & Seyfert 1 &2649&36 &0.05\\
NGC 3521 & Sbc    & normal    &805&8  &0.05\\
NGC 4051 & Sbc    & Seyfert 1 &725&8  &0.02\\
NGC 4278 & E1     & LINER     &649&9.7  &0.02\\
NGC 5033 & Sc     & LINER     &875&12  &0.02\\
Mrk 3    & S0     & Seyfert 2 &4050&56 &0.13\\
Mrk 620  & SBa    & Seyfert 2 &1840&24 &0.09\\
He 2-10 &dwarf Irr&starburst  &873&8  &0.15\\
MCG 8-11-11& SB   & Seyfert 1 &6141&80 &0.32\\
M 81     & Sb     & LINER     &-30&3.5  &0.06\\
\noalign{\smallskip}
\hline
\end{tabular}
\end{table*}

The instrument used was the Herzberg long-slit spectrograph equipped
with a 512x512 pixel CCD camera with a spatial sampling of 0.57''/px.
A grating of 300 lines/mm with a dispersion of 3.3\AA\ per pixel was
used. The spectral resolution is 8.5\AA\ FWHM for a 1.2'' slit width.
The seeing ranged between 1'' and 1.5''. The slit, which has a length
of 130 arcsec, was always set parallel to the paralactic angle.

The data are 2-dimensional (2D) spectra in three different
wavelength ranges: 5005-6640\AA, 6585-8225\AA\ and 8140-9785\AA, with
a small overlap which allows us to merge the three regions together.

The data reduction was carried out using the ESO-MIDAS package.
Standard reduction procedures for bias subtraction and flat-fielding
were applied.  Rebinning flux onto a linear wavelength scale was
carried out row by row, using the spectrum of an iron/argon
calibration lamp observed before and after each exposure. The data
were then corrected for airmass using the extinction curve provided in
the CFHT Observer's manual. Finally, flux calibration was performed in
the three wavelength ranges separately. The spectrophotometric
standard stars, BD+8~2015 (Stone, 1977), HD~84937 and HD~19445 (Oke
and Gunn, 1983) were observed at intervals throughout each night.  The
response function of the CCD was found to be uniform along the slit,
so an average calibration curve was derived and applied to each row
individually. A weighted mean calibration curve for each wavelength
domain was constructed for each night.

Sky subtraction was then performed. By measuring the sky levels in
frames of compact galaxies and stars, we determined that the sky level
was uniform, i.e. no gradient was observed except for the regions
closer than 40 arcsec to the edges of the frames. Therefore by
restricting our analysis to data in the central 50 to 100 arcsec of
the slit, no correction for sensitivity gradients along the slit need
be applied. The final sky spectrum was thus the average of a number of
cross-sections on both sides of the galaxies, not too far from the
sampled regions. However, some of the galaxies studied have optical
images extending over a few arcminutes, and hence starlight is present
over the full length of the slit. But, since our integration times are
relatively short and since the light profile drops rapidly, the
stellar contribution from the outer regions of the galaxy is
negligible. This has been confirmed by comparison of the sky spectra
away from the nucleus of extended galaxies, with those of more compact
galaxies. These spectra are effectively the same,
implying that we do not detect starlight far away from the nucleus in
any galaxy of our sample apart for M~81 which has an effective radius
of 200 arcsec. In that case it was verified that absorption line
equivalent width measurements, performed in spectra extracted
$\sim$30 arcsec from the center, had the same value when sky
subtraction was done using the edges of the frame than when a sky
spectrum built from frames of compact galaxies was used. The only
effect of weak sky contamination by the galaxy is to give final sky
subtracted spectra with somewhat larger a S/N ratio.

Logarithmic flux profiles across the galaxy 2D spectra were used to
determine where to extract the 1D spectra of the nucleus and the
regions around it. For the nucleus we have summed between 4 and 6
rows, and for the surrounding regions between 4 and 8, depending on
the seeing, the extent of the object and the S/N required. When the
spectra of two symmetric regions lying on either side of the nucleus
were very similar, indicating that the stellar population of the
galaxy is homogeneous, we averaged them together to provide a mean
spectrum.  Henceforth this mean spectrum will be referred to as {\it
  the ring}, although it is long-slit data, in order to distinguish it
from single region spectra.  Some of the galaxies were not spatially
resolved, and for these only the nuclear spectrum is presented. For
nearby bright galaxies we could select up to 3 regions on each side of
the nucleus.  Table~2 gives the position angle of the slit, the
effective radius (RC3) and the size of the regions defined along the
slit for each galaxy: the fourth column gives the diameter of the
nuclear region and the following columns give, on the first line the
distance from the centre, and on the second line the extension, in
parsecs, of each extranuclear region.

\begin{table*}
\caption[]{The regions defined in each galaxy}
\bigskip
\label{espectros}
\begin{tabular}{|l|r|l|r|c|c|c|}
\hline
\noalign{\smallskip}
\multicolumn{1}{c}{galaxy}&\multicolumn{1}{c}{PA}&\multicolumn{1}{c}{R$_e$}&\multicolumn{1}{c}{nucleus}&reg1a/reg1b&reg2a/reg2b&reg3a/reg3b\\
 &\multicolumn{1}{c}{($^o$)}&\multicolumn{1}{c}{(kpc)}&\multicolumn{1}{c}{(pc)}&(pc)&(pc)&(pc)\\
\noalign{\smallskip}
\hline
\noalign{\smallskip}

NGC 1275& -30&5.9 &1200&1100/1100&      - & - \\
        &    & &    &1000/1000&        &   \\
NGC 2110& -25&- &520& 440(ring)&   830/-& - \\
        &    & &   & 350      &   430  &   \\
NGC 3227& -85&2.7 &330&        - &      - & - \\
        &    & &   &          &        &   \\
NGC 3310& -20&0.7 &310&  360/310 & 740/620& - \\
        &    & &   &  410/310 & 360/310&   \\
NGC 3379&  55&2.2 &110& 110(ring)&230(ring)&360(ring)\\
        &    & &   &  110     & 130    & 130    \\
NGC 3516&  10&2.0 &600&  500/500 &      - & - \\
        &    & &   &  400     &        &   \\
NGC 3521&  40&2.9 & 400 & - &   - & - \\
        &    & &   & - & -  & - \\
NGC 4051&  40&2.7 &240&        - &      - & - \\
        &    & &   &          &        &   \\
NGC 4278& -65&1.0 &260& 200(ring)& 330(ring)& - \\
        &    & &   & 130      & 130      &   \\
NGC 5033&  30&4.8 &230&     -/210&   -/420& - \\
        &    & &   &       190&     230&   \\
Mrk 3   & -10&1.4 &1070&1000(ring)&     - & - \\
        &    & &    & 930      &       &   \\
Mrk 620 &  10&3.0 &460& 420(ring)&      - & - \\
        &    & &   & 400      &        &   \\
He 2-10 &  20&- &180&      -   &      - & - \\
        &    & &   &          &        &   \\
MCG 8-11-11&-30&- &2600&        - &      - & - \\
        &    & &   &          &        &   \\
M 81    &  25&3.5 & 70&  70(ring)&160(ring)&250(ring) \\
        &    & &   &    80    &     100 &  100  \\
\noalign{\smallskip}
\hline
\end{tabular}
\end{table*}

Finally, as we have no absolute flux calibration because the nights were
not photometric, the median flux ratio in the overlapping region was
used to normalize one segment to the next, resulting in a spectrum
covering the range 5000-9780\AA.

The atmospheric correction was performed with special attention, as in
this wavelength range there are many strong atmospheric bands of
H$_2$O and O$_2$ which affect the metallic lines of interest for
stellar population synthesis. O and B stars were observed for this
purpose since they are reasonably featureless. Ideally, if one wishes
to obtain good atmospheric band cancellation, one should observe an
early-type star at the same airmass and quasi-simultaneously to the
programme objects, since atmospheric absorption is a function of time
and zenith distance.  Numerous stars were observed but for practical
reasons not generally at a similar airmass to the galaxies.  Therefore
one weighted mean and normalized atmospheric spectrum was created. The
identification of the bands are mainly based on the work of Pierce \&
Breckinridge (1973), Vreux et al. (1983) and Wade \& Horne (1988).
Comparison of the strength of the O$_2$ and H$_2$O bands in the
individual spectra of the atmospheric standards showed that the
relative contribution of O$_2$ and H$_2$O vary during each night.
We therefore corrected separately for O$_2$ and H$_2$O. The 1D galaxy
spectra were divided by $[1-(1-atm)*K]$ where $atm$ is the atmospheric
O$_2$ or H$_2$O band spectrum and $K$ is a constant usually equal to
1, but which differs from unity whenever $atm$ is deeper or smoother
than the atmospheric bands present in the galaxy.

Figure 1 (a to o) shows the spectra extracted from the 15 galaxy
frames, corrected for galactic interstellar reddening using the
galactic reddening law from Howarth (1983). The spectra of the
surrounding regions are displayed with individual scaling factors
chosen in order to allow the best comparison of the stellar population
properties of the different regions. All spectra are displayed at rest
wavelength.

Equivalent width (EW) of about 40 stellar features have been measured
in almost every spectrum with the aim of performing population
synthesis (see paper II). The nuclei of the Seyfert 1's and of Mrk~3,
a Seyfert~2, have not been measured as they are largely dominated by
emission lines. The spectra with too low signal to noise ratio to
allow valuable synthesis were omitted as well. Identification and
wavelength range of the absorption lines and bands, together with the
measured EW can be found in the Appendix (Table A1 to 12). The ranges
have been defined taking into account the shape of the absorption
features in the spectra of both hot and cold stars in the stellar
library. The line EW were measured using a continuum level defined
over the entire wavelength range available.  This method results in a
more reliable determination of the line strength, compared to other
methods wich use only a locally defined continuum.  Note that in
stellar population synthesis the use of a {\it global} continuum is
crucial, since all spectral features play a role in determining the
best fit composite spectrum.

\begin{figure*}
\begin{center}
\psfig{figure=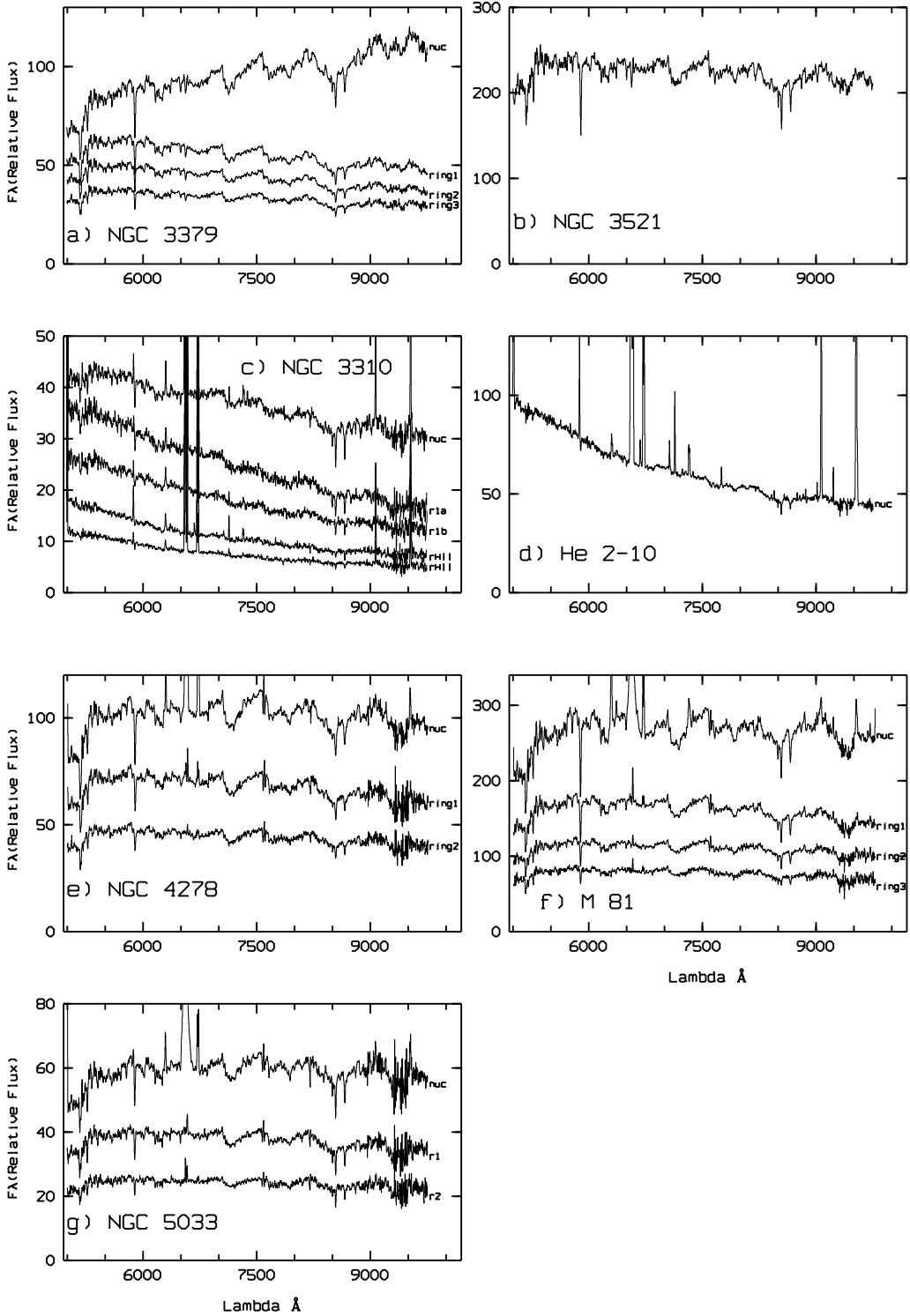,height=21cm}
\end{center}
\caption{Spectra observed in the central region of a) NGC~3379. For
clarity, the spectra of the off-nuclear regions have been multiplied
by a scaling factor of 1.2, 2.0, 2.8, for ring 1, 2 and 3, respectively;
b) Similarly for NGC~3521; c) NGC~3310: scaling factor of 2.0, 1.6,
1.0 and 1.0, for region 1a, 1b, 2a and 2b, respectively; d) He~2-10; 2)
NGC~4278: scaling factor of 5.0 and 6.5, for ring 1 and 2,
respectively; f) M~81: scaling factor of 1.7, 2.8, 4.2, for
ring 1, 2, 3, respectively; g) NGC~5033: scaling factor of
2.2 and 3.5, for region 1 and 2, respectively.}
\end{figure*}

\addtocounter{figure}{-1}
\begin{figure*}
\begin{center}
\psfig{figure=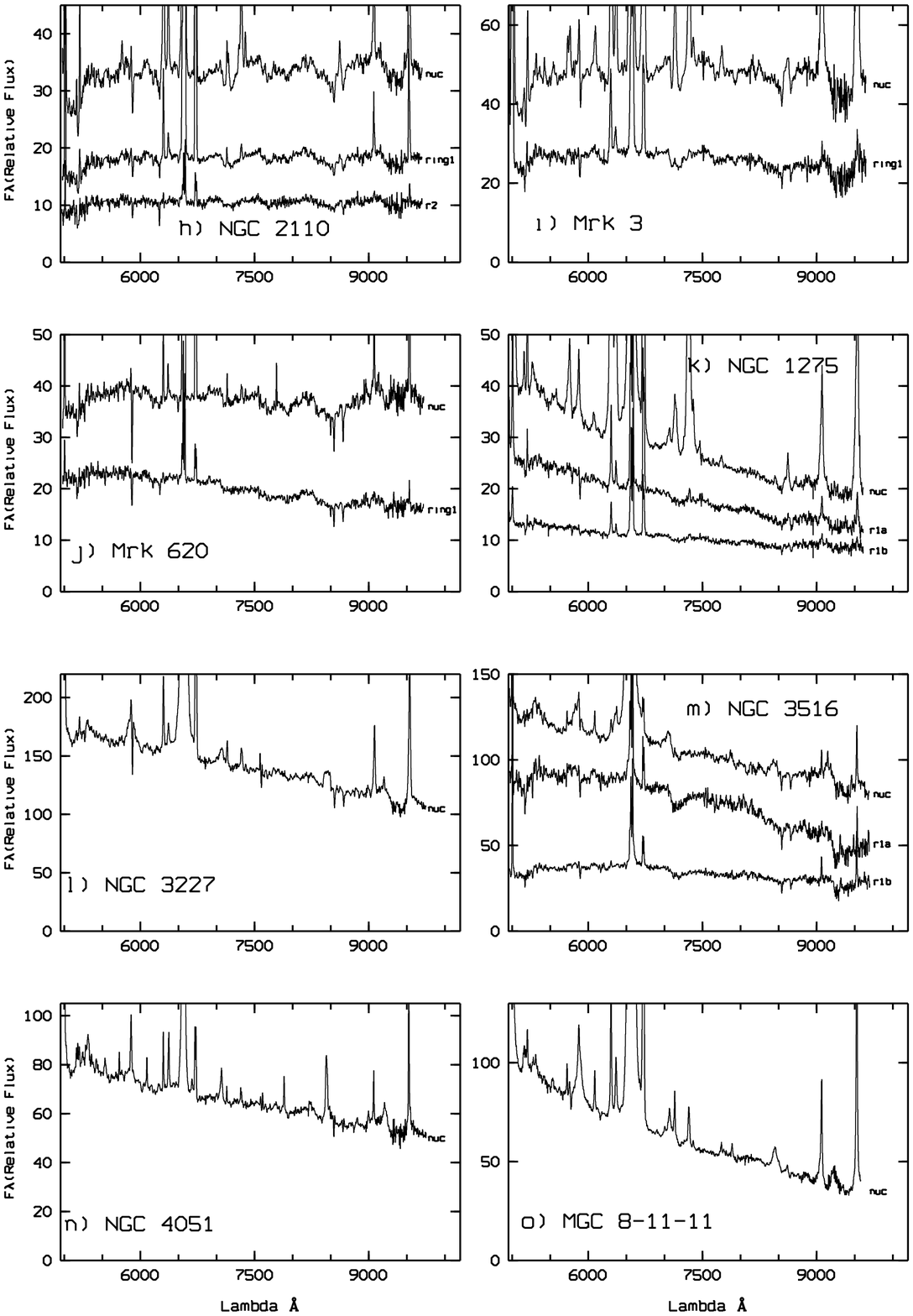,height=21cm}
\end{center}
\caption{Cont. h) NGC~2110:
scaling factor of 2.0  for ring 1 and region 2; i) Mrk~3:
scaling factor of 2.8 for ring1; j) Mrk~620: scaling
factors of 2.5 for ring 1; k) NGC~1275: scaling factors of 2.6 and 1.7,
for regions 1a and 1b, respectively; l) NGC~3227; m) NGC~3516:
scaling factors of 3.5; n) NGC~4051; o) MCG~8-11-11}
\end{figure*}

\section{The sample}

\subsection{The template galaxies}
     
The non-active galaxies in the sample are NGC~3379 (Fig. 1a) and
NGC~3521 (Fig. 1b), a standard giant elliptical and a spiral galaxy
respectively. No emission lines are present in any of the two
galaxies. The classification of NGC~3521 as a LINER by some authors
(e.g. Spinoglio \& Malkan, 1989) is not supported by our data.
     
The nuclear spectrum of NGC~3379 is very red with strong molecular
features (MgH, TiO, CN) implying a dominant cool population and
significant internal reddening due to the presence of dust. The
extranuclear regions of NGC~3379 (Rings 1, 2 and 3) are much bluer
than the nucleus, although the molecular features are also very
strong. Our flux ratios for the three off-nuclear rings are in fair
agreement with the V (Goudfrooij et al., 1994) and R (Peletier et al.,
1990) surface brigthness gradient. The very red nucleus (inner 3'')
can be seen in the colour gradients, (B-I) and (V-I) from Goudfrooij
et al. (1994).  This colour gradient (E$_{B-V}$=0.26) could be
intrinsic to the stellar population or alternatively it could result
from the presence of centrally localised dust. The fact that a dust
lane (van Dokkum and Franx, 1995) has been detected, and that the
spectra are similar when regions are compared to the derredened
nucleus, favours the latter hypothesis.  Population gradients have
been extensively studied in the central part of this elliptical
galaxy. All authors agree that spectral gradients exist, but they do
not agree on the slope (Davidge, 1992; Davies et al., 1993; Delisle \&
Hardy, 1992). In any case, gradients are steep only beyond 6 arcsec
from the nucleus. Although EW are not indices a rough comparison of
our measurement to published gradients can be made. The two first
lines in Table A1 (labelled FeI and MgI+MgH) can be compared to the
Mg$_2$ index. If one excepts the red nucleus, that is the central 3
arcsec, only Region 3 (located beyond 6 arcsec from the nucleus) has a
lower EW compared to regions 1 and 2. That is consistent with the
gradient published by Davies et al. (1993). Even though
FeI$\lambda$5270 does not show gradient some other blends of FeI, CN,
CaI and VO do. This again is consistent with the marginal variation of
FeI$\lambda$5270 index found by Davies et al. between 3 and 9 arcsec
from the center. However most absorption line EW in Table A1 display
no variation from one region to the other one, and from Fig. 1a it can
be seen that the circum-nuclear regions, extending from 100 to 400 pc,
have remarkably similar spectra implying a very homogeneous central
part.
     
  A problem with the telescope focus, while observing NGC~3521,
 resulted in a defocused image and therefore precluded the
 possibility of obtaining spatially resolved spectra in this case.  
 Only a spatially averaged region corresponding to 400pc in diameter
 could be extracted. This spectrum is similar to that extracted within
 the inner 10 arcsec for NGC~3379. To our knowledge no published
 information on colour gradients is available for NGC~3521.

\subsection{The starburst galaxies}
     
     Spectra of two starburst galaxies, NGC~3310, a spiral, and
     He~2-10, a dwarf irregular, are shown in Figures 1c and d.
     
     Compared to the normal galaxies, the nuclear spectra of the
     starbursts are much bluer (although the internal reddening can
     still be high; e.g. Hutsemekers \& Surdej, 1984; Grothues \&
     Schmidt-Kaler, 1991) and the cool old stellar component is weaker
     (in particular the TiO bands). The strong emission lines
     characteristic of HII regions are present.
 
     No circum-nuclear region was extracted from around the nucleus of
     the dwarf galaxy He~2-10 because of its low surface brightness.
     Johansson (1987) estimated that $\simeq 1\%$ of the luminous mass
     inside 5 arcsec (150pc) of He~2-10 is in the form of newly formed
     stars. The emission line blend characteristic of Wolf-Rayet stars
     has been observed in the nucleus by Hutsemekers \& Surdej (1984).
     In such galaxies the current rate of star formation is much
     higher than the average past rate. The metallicity is low (e.g.
     Masegosa et al., 1994; Marconi et al., 1994; Walsh \& Roy, 1993).
     
     In NGC~3310 very luminous HII regions are observed within 20
     arcsec of the nucleus. These have moderately low metal abundance
     compared to the inner regions of galaxies of a similar Hubble
     type (Heckman \& Balick, 1980, Pastoriza et al., 1993).  This
     effect could be a consequence of a merger with a low metallicity
     galaxy (Balick \& Heckman, 1981; Grothues \& Schmidt-Kaler,
     1991). The nucleus itself has a solar abundance (Pastoriza et
     al., 1993).  The circum-nuclear region of NGC~3310, located
     around 330pc from the centre (Region 1a and 1b), is even bluer
     than the nucleus (E$_{B-V}$=0.25). When the continua are scaled
     up to the same level, it can be seen that the stellar spectra are
     identical on both sides of the nucleus showing a weak cool
     stellar component and the same emission lines are present in the
     nucleus. The two farther regions (Region 2a and 2b) are
     contaminated by HII regions located at 740 and 620pc from the
     centre (see Fig. 1 in Pastoriza et al., 1993).  An increase of
     the emission line strengths in Region 2a is observed with respect
     to the bulge component as it includes a significant contribution
     from the HII region, while Region 2b, close to another HII
     region, has emission line equivalent widths reduced by the bulge
     stellar continuum. It is clear that in NGC~3310 an intense phase
     of star formation is occuring over an extended region, and that
     this dominates over the cool stellar component.

\subsection{The LINERs}

The LINERS in our study are the Sb and Sc
galaxies, M~81 (Fig. 1f) and NGC~5033 (Fig. 1g) respectively, as well
as the giant elliptical NGC~4278 (Fig. 1e).
NGC~5033 is sometimes classified as a Seyfert 1.9 galaxy, while
Filippenko \& Sargent (1985) state that its emission lines have some
properties of both LINERs and Seyferts.  Koratkar et al. (1995) detect
a broad component in H$\alpha$ and H$\beta$ but its strength strongly
depends on the estimated continuum level and on the subtraction of the
stellar continuum. A spectral comparison with the other LINERs and
Seyfert 1s presented here shows that the spectrum of NGC~5033
(observed in January 1990) does not exhibit noticeable Seyfert
characteristics in the optical range. All three LINERs exhibit quite
intense narrow emission lines in their nuclei plus weak H$\alpha$
emission in the surrounding regions.

A strong contribution from cool stars is obvious in all three
galaxies. Circum-nuclear stellar populations appear quite homogeneous,
that is that one can scale any region to the other one without
striking difference. However, the nuclei are somewhat redder. This
difference can be explained either by dust reddening in the nucleus
(E$_{B-V}$ $\sim$ 0.07) or by the presence of a blue stellar component
in the circum-nuclear region.

Near-IR line gradients are observed in M81 (Delisle \& Hardy, 1992)
but we cannot easily compare with results in Table A6, as CaIIT
and TiO have been measured in a different way both in terms of
continuum and blend of lines. Colour gradients similar to what we
find, the nuclear starlight being redder than the bulge some 20 arcsec
away, are quoted in Keel (1989). Nevertheless, a characteristic of our
spectra of M81 is the strong similarity between the different regions
of the bulge (Rings 1, 2 and 3).

Colour and line index gradients have been observed in NGC~4278 by,
respectively, M$\o$ller et al. (1995), and Davidge \& Clark (1994) and
Davies et al. (1993). Albeit we do find similar variations in the blue
(below 5500\AA), no strong gradient is apparent on the overall
spectra. The change in continuum slope between the nucleus and
circumnuclear regions can again be attributed to either nuclear dust
or an additional small young component in the circum-nuclear regions.

\subsection{The Seyfert 2 galaxies}

Three Seyfert 2 galaxies are included in the sample: two early-type
galaxies, NGC~2110 and Mrk~3 and an SBa, Mrk~620. Note that NGC~2110
and Mrk~3 have obscuring material close to their centres, and are
suspected to hide a Seyfert 1 nucleus (Turner \& Pounds, 1989; Miller
\& Goodrich, 1990; Tran, 1995).
     
The continuum spectra of these three Seyfert 2 nuclei (Fig. 1h, 1i and 1j)
are flat and do not differ markedly from those of the LINERs, having 
strong absorption features from an integrated cool star spectrum on
which are superimposed the high ionization emission lines characteristic
of Seyfert 2s.
     
A circumnuclear ring spectrum at a galactocentric distance of $\sim$
450 pc plus another from an external region at $\sim$ 850 pc have been
extracted from the bulge spectrum of NGC~2110. When scaled, the
stellar populations of these regions appear similar to that of the
nucleus. The same situation is found for the spectra of the
circumnuclear regions of Mrk~620 and Mrk~3, although the circumnuclear
continuum shapes of these two galaxies are rather more blue than those
of their nuclei (E$_{B-V}$=0.16 and 0.08 respectively). A contribution
from the extended narrow line region (ENLR) is conspicuous in all
these extranuclear regions.

\subsection{The Seyfert 1 galaxies}
 
Five Seyfert 1 galaxies, NGC~1275 (Fig. 1k), NGC~3227 (Fig. 1l),
NGC~3516 (Fig. 1m), NGC~4051 (Fig. 1n) and MCG~8-11-11 (Fig. 1o) are
included in our sample. Only for NGC~3516, an SB0 galaxy, and
NGC~1275, a cD, was it possible to extract a spectrum from the
circumnuclear region; for the three other objects only the nuclear
spectrum has sufficient S/N ratio. Although sometimes classified as a
BL~Lac type AGN, NGC~1275 has a nuclear spectrum similar to that of a
Seyfert 1 in the wavelength range 5000-9800\AA.

The five Seyfert 1 nuclei have broad emission lines superposed on a
strong blue continuum spectrum. The stellar component, although detectable,
is much weaker than in the other types of galaxy in our sample.
This has been interpreted as due to dilution of the nuclear stellar
component by an additional featureless component, the nature of which
is still the subject of debate (e.g. Malkan \& Filippenko, 1983;
Terlevich et al., 1990). 
     
Serote Roos et al. (1996) have shown that the bulge of NGC~3516 is
quite homogeneous with a cool stellar population similar to that found
in the bulges of normal galaxies. In this paper the extent of the
off-nuclear regions has been defined somewhat differently, having a
smaller circum-nuclear region extracted from the 2D frame (400pc wide
instead of 600pc). The extranuclear spectra extracted on either side
of the nucleus were not averaged together, since one is bluer than the
other.  Nevertheless the conclusions drawn are essentially the same as
those of Serote Roos et al. as these modifications are relatively
minor.
     
The off-nuclear spectrum of NGC~1275 is very different from that of NGC~3516
having a bluer continuum and a moderately strong cool stellar
component similar to that found in starburst galaxies.  Malkan and Filippenko
(1983) also found that the host galaxy of NGC~1275 is bluer than
normal galaxies in agreement with Minkowski (1968) who first suggested
the existence of a population of young stars.
     
In fact using the HST WFPC, Holtzman et al. (1992) discovered a
population of young star clusters lying within 5kpc of the nucleus. 
Ferruit and P\'econtal (1994) obtained 2D spectra of the
central 10'' (3.5 kpc) of NGC~1275 and detected emission lines 
associated with some of these star clusters. They interpreted this as
evidence for continuing star formation as matter is accreted from the
cooling flow. Our circum-nuclear spectra cover some of these young
star clusters.
     
     In NGC~3227 an HII type emission spectrum extending over a region of about
     300pc from the nucleus suggests the presence of an extranuclear
     starburst (Arribas \& Mediavilla, 1994). 

\section{Discussion}

  The contribution to the total spectral energy
distribution (SED) of AGN arising from the stellar component of the underlying
host galaxy, can be estimated from the strengths of absorption line strengths, 
by comparison with the corresponding strengths found in normal galaxies.
A limitation of this
method arises from the fact that {\it normal} galaxies have internal
reddening, velocity dispersion, central stellar metallicity and
population which may not necessarily be an appropriate match to the
underlying stellar component at the centre of an active galaxy.

It is possible to alleviate some of the above mentioned problems
by comparing the stellar absorption line strengths in the nucleus,
with those in the surrounding regions of the bulge for the {\it same}
galaxy, however this requires that no stellar population or dust
gradient be present. We should point out that an
excess of hot stars has the same effect on the shape of the spectra as
would have a small reddening correction. When gradients are
obviously present, a population synthesis of the different regions
using data of the highest possible  spatial resolution is needed. 

In the nuclei of the Seyfert~1s, as expected, the contribution of the
stellar component is not the dominant feature, only the CaIIT lines
are conspicuous. We scaled the circumnuclear stellar population in
both NGC~3516 and NGC~1275, to that of the nucleus, using the
equivalent width ratio of a reference line, CaII~8542\AA\ (cf. Serote
Roos et al., 1996a). The equivalent widths were measured using a
continuum level defined over the whole wavelength range. This method
allows a better determination of the nucleus to bulge line ratio as it
takes into account both the contribution of CaIIT and TiO, avoiding
underestimation of the EWs in the cases where cool stars dominate the
stellar population. In both Seyfert 1's, 65--85\% for NGC~1275 and
50--65\% for NGC~3516 of the continuum emission around CaIIT can be
attributed to the host galaxy as long as one suppose that no gradients
of any type is present. In those two cases a non-stellar featureless
continuum can be inferred as one might expect for Seyfert 1s.

It should be added that, although gradients within a galaxy and/or
very different populations from object to object are found, the global
properties of the host galaxies are similar when large aperture are
used (say about 10 arcsec for nearby galaxies). If one extracts the
spectral energy distribution (SED) within 10 arcsec for the
elliptical galaxies NGC~2110 and NGC~3379, and for the Sc galaxies,
NGC~5033 and NGC~3521, there is a relationship between SED and
morphological type. This is not surprising as we are studying the
central regions/bulges of these galaxies and in no case are we
sampling the disk. It is well established that ellipticals and the
bulges of spirals have very similar optical SEDs. However, in our
study we are not concerned with their global properties but rather in
disentangling the nuclear component from the stellar features of the
host galaxy. In this context we need good spatial resolution in order
to isolate the unresolved nucleus, particularly since the effect we
are looking for may be weak, apart from Seyfert 1s, and will be
diluted within large apertures.  On smaller scales the gradients can
be important e.g. if one compares the two Sc galaxies, NGC~3310 and
NGC~5033, it is obvious that they do not have the same properties.
Line strengths, as e.g. MgH band or CaIIT, are not identical and
therefore spurious dilutions would be deduced from direct comparison.
This is partially due to very different effective radii, which must be
taken into account.

\section{Conclusions}

 The nuclei of our sample galaxies are generally redder than the outer
regions (this excludes Seyfert 1s as their nuclear
spectra are dominated by strong broad emission lines).  This colour
gradient could either be due to dust or to stellar population
gradients. In addition to dust/population gradients, the presence of a
featureless continuum is inferred in a number of AGN.  This component
might be of non-thermal origin, plausible in Seyfert galaxies,
or of stellar origin in the case of nuclear starburst activity. The
extra component dilutes the strengths of the stellar absorption lines in the
spectra of the nuclear regions of AGN. A detailed study of the spectral
shape of the diluting component will allow us to differentiate between
the two hypotheses.

A full population synthesis analysis for most of the galaxies in our
sample, using a new method developed to determine unique solutions, is
presented in Paper II.

\section*{Acknowledgments}

M. Serote Roos acknowledges financial support from JNICT, Portugal,
under grant no. PRAXIS XXI/BD/5270/95.
We thank M. Caillat for the writing of the MEASURE program that we
used to estimate the equivalent widths.

\newpage

\appendix
\section{ Absorption line measurements}
\label{ap.1}

The equivalent widths (EW) presented here were measured with respect
to a global continuum (see section 2). Positionning the continuum is
however somewhat subjective as stellar features can be heavily blended
and emission lines can be present. Consequently the errors on the EW
are not simple to evaluate. One should consider both the signal to
noise ratio of individual spectra and EW differences between the
lowest and the highest acceptable continuum level. The latter
uncertainty amounts to 0.4\AA\ for most features. However, it can vary
up to 1\AA\ in the case of shallow features as CN. Actually this error
is dominating over all other measurement and statistical
uncertainties.

Column 1 of Table A1 to 12 gives the line identification of the
features. Column 2 displays the wavelength ranges used to measure the
EW. The EW measured in the nucleus and the circum-nuclear regions are
given in the following columns. Whenever the emission lines dominate
a measured wavelength range the value of the EW is omitted from the
table.

\begin{table}
{\scriptsize
\caption{NGC 3379}
\centerline{
\begin{tabular}{|l|c|c|c|c|c|}
\noalign{\smallskip}
\hline
\noalign{\smallskip}
identification& $\lambda$ (\AA)&nucleus&ring 1&ring 2&ring 3\\
\noalign{\smallskip}
\hline
\noalign{\smallskip}
FeI&5058-5156             &21.7&19.0&19.2&18.8\\
FeI,MgI+MgH&5156-5240     &20.3&18.5&18.3&17.7\\
FeI&5240-5308             &7.2&6.5&6.3&6.5\\
FeI&5308-5356             &2.5&2.2&2.6&2.6\\
FeI&5356-5421             &3.4&3.0&3.3&3.3\\
FeI,TiO,MgI&5421-5554     &7.8&7.2&8.4&7.8\\
CaI,FeI,TiO&5554-5630     &4.6&4.9&5.3&4.5\\
FeI,TiO&5630-5676         &2.1&2.1&2.1&1.8\\
FeI,NaI&5676-5725         &1.5&1.9&1.9&1.4\\
FeI,TiO&5725-5825         &0.9&2.1&2.5&2.1\\
FeI,TiO,CaI&5825-5874     &0.6&1.5&1.4&1.0\\
NaI&5874-5914             &6.3&6.5&5.7&5.6\\
FeI,Ti,MnI&5914-6029      &4.1&5.6&5.7&5.5\\
FeI,CaI&6029-6110         &0.4&1.6&1.8&2.0\\
FeI&6110-6148             &0.9&1.7&1.7&1.5\\
TiO,CaI&6148-6208         &4.7&5.4&5.7&5.3\\
TiO,FeI&6208-6325         &9.3&10.3&10.3&10.2\\
FeI,CaH,TiO&6325-6374     &2.7&3.1&3.0&2.9\\
FeI,CaI&6374-6481         &2.4&3.3&3.4&2.4\\
FeI,TiO,CaI,BaII&6481-6535&2.8&2.8&3.1&3.1\\
$H_{\alpha}$,TiO&6535-6582&2.4&3.4&3.6&3.6\\
FeI,TiO&6582-6622         &1.7&2.1&2.2&2.2\\
TiO&6622-6651             &1.7&2.0&1.9&2.0\\
FeI,TiO&6651-6690         &2.8&3.0&3.1&3.0\\
FeI,TiO,CaI,Si&6690-6761  &5.2&6.3&5.8&5.6\\
FeI,TiO,CaH&6761-6795     &2.1&2.6&2.8&2.6\\
TiO,CaH&6795-6827         &1.8&2.5&2.3&2.1\\
FeI,SiI&6969-7048         &0.7&1.6&1.6&1.4\\
TiO&7048-7078             &1.6&2.1&2.1&1.7\\
TiO,NiI,FeI&7078-7140     &7.2&7.9&7.5&7.1\\
TiO,VO&7341-7377          &1.0&1.6&1.3&1.3\\
FeI,VO&7377-7434          &0.9&1.9&1.9&1.8\\
FeI,VO&7434-7482          &0.0&0.8&0.0&0.2\\
FeI,VO&7482-7540          &0.0&0.2&0.0&0.0\\
TiO,OI&7737-7823          &5.3&7.0&6.5&6.8\\
VO,CN,FeI&7823-7888       &4.0&5.4&5.2&4.6\\
VO,CN,FeI&7888-7970       &7.5&8.7&8.0&8.1\\
FeI,CN&7970-8060          &4.2&5.5&4.3&4.6\\
TiO,FeI&8411-8453         &5.3&5.7&5.6&5.7\\
TiO,FeI&8453-8490         &5.6&5.9&5.7&5.8\\
CaII&8490-8508            &4.0&3.6&3.5&4.1\\
FeI,VO,TiI&8508-8527      &3.2&3.9&3.9&3.4\\
CaII&8527-8557            &7.6&7.8&7.3&7.2\\
FeI,VO,TiO&8557-8645      &10.6&11.0&11.8&12.1\\
CaII&8645-8677            &6.0&6.2&6.2&6.3\\
FeI,MgI&8677-8768         &9.2&9.9&9.9&11.1\\
FeI,MgI&8768-8855         &5.1&6.2&6.9&7.1\\
\noalign{\smallskip}
\hline
\end{tabular}
}
}
\end{table}

\begin{table}
{\scriptsize
\caption{NGC 3521}
\centerline{
\begin{tabular}{|l|c|c|}
\noalign{\smallskip}
\hline
\noalign{\smallskip}
identification& $\lambda$ (\AA)& total\\
\noalign{\smallskip}
\hline
\noalign{\smallskip}
FeI&5058-5156&15.3\\
FeI,MgI+MgH&5156-5240&15.4\\
FeI&5240-5308&6.4\\
FeI&5308-5356&1.9\\
FeI&5356-5421&2.4\\
FeI,TiO,MgI&5421-5554&5.9\\
CaI,FeI,TiO&5554-5630&4.1\\
FeI,TiO&5630-5676&2.6\\
FeI,NaI&5676-5725&2.2\\
FeI,TiO&5725-5825&2.7\\
FeI,TiO,CaI&5825-5874&1.2\\
NaI&5874-5914&7.9\\
FeI,Ti,MnI&5914-6029&5.9\\
FeI,CaI&6029-6110&2.0\\
FeI&6110-6148&1.8\\
TiO,CaI&6148-6208&5.3\\
TiO,FeI&6208-6325&9.6\\
FeI,CaH,TiO&6325-6374&3.2\\
FeI,CaI&6374-6481&2.7\\
FeI,TiO,CaI,BaII&6481-6535&3.4\\
$H_{\alpha}$,TiO&6535-6582&3.7\\
FeI,TiO&6582-6622&2.0\\
TiO&6622-6651&1.5\\
FeI,TiO&6651-6690&2.6\\
FeI,TiO,CaI,Si&6690-6761&4.7\\
FeI,TiO,CaH&6761-6795&1.7\\
TiO,CaH&6795-6827&2.0\\
FeI,SiI&6969-7048&1.8\\
TiO&7048-7078&1.4\\
TiO,NiI,FeI&7078-7140&6.7\\
TiO,VO&7341-7377&1.4\\
FeI,VO&7377-7434&2.0\\
FeI,VO&7434-7482&0.8\\
FeI,VO&7482-7540&0.4\\
TiO,OI&7737-7823&4.5\\
VO,CN,FeI&7823-7888&3.2\\
VO,CN,FeI&7888-7970&6.7\\
FeI,CN&7970-8060&3.8\\
TiO,FeI&8411-8453&5.1\\
TiO,FeI&8453-8490&4.8\\
CaII&8490-8508&3.3\\
FeI,VO,TiI&8508-8527&3.5\\
CaII&8527-8557&7.2\\
FeI,VO,TiO&8557-8645&9.7\\
CaII&8645-8677&5.7\\
FeI,MgI&8677-8768&8.9\\
FeI,MgI&8768-8855&5.0\\
\noalign{\smallskip}
\hline
\end{tabular}
}
}
\end{table}

\begin{table}
{\scriptsize
\caption{NGC 3310}
\centerline{
\begin{tabular}{|l|c|c|c|c|}
\noalign{\smallskip}
\hline
\noalign{\smallskip}
identification& $\lambda$ (\AA)&nucleus&region 1a &region 1b\\
\noalign{\smallskip}
\hline
\noalign{\smallskip}
FeI&5058-5156&5.8&3.6&4.2\\
FeI,MgI+MgH&5156-5240&6.3&6.2&5.5\\
FeI&5240-5308&2.5&1.8&2.3\\
FeI&5308-5356&0.5&0.6&1.3\\
FeI&5356-5421&0.1&0.5&0.0\\
FeI,TiO,MgI&5421-5554&0.9&3.3&2.3\\
CaI,FeI,TiO&5554-5630&1.0&2.3&3.7\\
FeI,TiO&5630-5676&0.8&1.4&2.4\\
FeI,NaI&5676-5725&0.0&1.3&1.6\\
FeI,TiO&5725-5825&-&2.2&2.7\\
FeI,TiO,CaI&5825-5874&-&0.6&1.0\\
NaI&5874-5914&1.7&2.5&3.1\\
FeI,Ti,MnI&5914-6029&3.0&6.3&6.3\\
FeI,CaI&6029-6110&1.9&3.4&3.5\\
FeI&6110-6148&1.5&2.6&2.7\\
TiO,CaI&6148-6208&4.3&5.4&5.1\\
TiO,FeI&6208-6325&-&-&-\\
FeI,CaH,TiO&6325-6374&2.0&3.0&2.4\\
FeI,CaI&6374-6481&3.9&6.1&4.7\\
FeI,TiO,CaI,BaII&6481-6535&2.8&3.6&3.2\\
$H_{\alpha}$,TiO&6535-6582&-&-&-\\
FeI,TiO&6582-6622&-&-&-\\
TiO&6622-6651&0.6&1.4&1.6\\
FeI,TiO&6651-6690&1.0&1.9&2.4\\
FeI,TiO,CaI,Si&6690-6761&-&-&-\\
FeI,TiO,CaH&6761-6795&0.7&1.3&2.2\\
TiO,CaH&6795-6827&0.5&2.2&2.1\\
FeI,SiI&6969-7048&-&0.8&2.8\\
TiO&7048-7078&-&0.9&1.2\\
TiO,NiI,FeI&7078-7140&1.5&3.7&3.6\\
TiO,VO&7341-7377&0.3&1.6&2.4\\
FeI,VO&7377-7434&0.5&1.5&2.3\\
FeI,VO&7434-7482&0.3&1.3&2.1\\
FeI,VO&7482-7540&0.1&0.8&1.2\\
TiO,OI&7737-7823&2.8&4.8&6.1\\
VO,CN,FeI&7823-7888&2.7&3.7&3.7\\
VO,CN,FeI&7888-7970&5.0&6.6&7.3\\
FeI,CN&7970-8060&2.7&2.5&6.2\\
TiO,FeI&8411-8453&3.6&4.0&4.2\\
TiO,FeI&8453-8490&3.7&4.1&4.2\\
CaII&8490-8508&3.1&3.3&2.0\\
FeI,VO,TiI&8508-8527&2.1&2.6&2.5\\
CaII&8527-8557&5.9&6.3&5.4\\
FeI,VO,TiO&8557-8645&5.8&8.8&8.4\\
CaII&8645-8677&5.0&5.7&5.1\\
FeI,MgI&8677-8768&5.3&7.2&7.6\\
FeI,MgI&8768-8855&2.3&7.3&6.1\\
\noalign{\smallskip}
\hline
\end{tabular}
}
}
\end{table}

\begin{table}
{\scriptsize
\caption{He 2-10}
\centerline{
\begin{tabular}{|l|c|c|}
\noalign{\smallskip}
\hline
\noalign{\smallskip}
identification& $\lambda$ (\AA)&nucleus\\
\noalign{\smallskip}
\hline
\noalign{\smallskip}
FeI&5058-5156&2.8\\
FeI,MgI+MgH&5156-5240&3.4\\
FeI&5240-5308&1.9\\
FeI&5308-5356&0.7\\
FeI&5356-5421&0.9\\
FeI,TiO,MgI&5421-5554&2.4\\
CaI,FeI,TiO&5554-5630&1.9\\
FeI,TiO&5630-5676&0.9\\
FeI,NaI&5676-5725&0.0\\
FeI,TiO&5725-5825&2.1\\
FeI,TiO,CaI&5825-5874&-\\
NaI&5874-5914&-\\
FeI,Ti,MnI&5914-6029&4.3\\
FeI,CaI&6029-6110&2.5\\
FeI&6110-6148&1.5\\
TiO,CaI&6148-6208&3.5\\
TiO,FeI&6208-6325&2.9\\
FeI,CaH,TiO&6325-6374&0.4\\
FeI,CaI&6374-6481&1.8\\
FeI,TiO,CaI,BaII&6481-6535&2.5\\
$H_{\alpha}$,TiO&6535-6582&-\\
FeI,TiO&6582-6622&-\\
TiO&6622-6651&-\\
FeI,TiO&6651-6690&-\\
FeI,TiO,CaI,Si&6690-6761&-\\
FeI,TiO,CaH&6761-6795&1.7\\
TiO,CaH&6795-6827&1.9\\
FeI,SiI&6969-7048&1.5\\
TiO&7048-7078&-\\
TiO,NiI,FeI&7078-7140&-\\
TiO,VO&7341-7377&1.2\\
FeI,VO&7377-7434&2.1\\
FeI,VO&7434-7482&1.8\\
FeI,VO&7482-7540&1.7\\
TiO,OI&7737-7823&-\\
VO,CN,FeI&7823-7888&2.8\\
VO,CN,FeI&7888-7970&5.2\\
FeI,CN&7970-8060&2.4\\
TiO,FeI&8411-8453&2.7\\
TiO,FeI&8453-8490&3.5\\
CaII&8490-8508&2.7\\
FeI,VO,TiI&8508-8527&2.1\\
CaII&8527-8557&5.1\\
FeI,VO,TiO&8557-8645&5.9\\
CaII&8645-8677&4.2\\
FeI,MgI&8677-8768&5.4\\
FeI,MgI&8768-8855&3.2\\
\noalign{\smallskip}
\hline
\end{tabular}
}
}
\end{table}

\begin{table}
{\scriptsize
\caption{NGC 4278}
\centerline{
\begin{tabular}{|l|c|c|c|c|}
\noalign{\smallskip}
\hline
\noalign{\smallskip}
identification& $\lambda$ (\AA)&nucleus&ring 1&ring 2\\
\noalign{\smallskip}
\hline
\noalign{\smallskip}
FeI&5058-5156&23.9&20.0&18.0\\
FeI,MgI+MgH&5156-5240&20.7&18.9&17.6\\
FeI&5240-5308&7.3&6.8&5.2\\
FeI&5308-5356&2.5&2.0&1.8\\
FeI&5356-5421&2.7&1.9&1.0\\
FeI,TiO,MgI&5421-5554&7.4&6.7&4.3\\
CaI,FeI,TiO&5554-5630&5.1&5.2&4.7\\
FeI,TiO&5630-5676&2.4&2.6&2.1\\
FeI,NaI&5676-5725&2.2&2.4&1.8\\
FeI,TiO&5725-5825&0.9&2.7&1.5\\
FeI,TiO,CaI&5825-5874&0.6&1.0&0.6\\
NaI&5874-5914&6.0&5.7&5.2\\
FeI,Ti,MnI&5914-6029&6.2&7.3&6.5\\
FeI,CaI&6029-6110&1.5&2.8&2.4\\
FeI&6110-6148&1.3&1.8&1.6\\
TiO,CaI&6148-6208&4.7&5.0&4.6\\
TiO,FeI&6208-6325&-&8.7&9.4\\
FeI,CaH,TiO&6325-6374&1.8&2.9&2.5\\
FeI,CaI&6374-6481&3.7&3.9&4.0\\
FeI,TiO,CaI,BaII&6481-6535&-&3.7&3.4\\
$H_{\alpha}$,TiO&6535-6582&-&-&2.3\\
FeI,TiO&6582-6622&-&-&1.8\\
TiO&6622-6651&1.9&1.9&2.3\\
FeI,TiO&6651-6690&2.9&3.2&3.6\\
FeI,TiO,CaI,Si&6690-6761&-&-&4.9\\
FeI,TiO,CaH&6761-6795&2.6&2.5&2.6\\
TiO,CaH&6795-6827&2.2&2.6&2.8\\
FeI,SiI&6969-7048&1.2&1.6&2.4\\
TiO&7048-7078&1.1&1.2&1.9\\
TiO,NiI,FeI&7078-7140&7.0&6.8&7.2\\
TiO,VO&7341-7377&1.4&1.7&1.5\\
FeI,VO&7377-7434&1.4&1.7&1.8\\
FeI,VO&7434-7482&0.5&1.0&0.0\\
FeI,VO&7482-7540&0.2&0.3&0.0\\
TiO,OI&7737-7823&6.4&6.9&6.8\\
VO,CN,FeI&7823-7888&4.9&4.7&4.9\\
VO,CN,FeI&7888-7970&8.2&7.7&7.5\\
FeI,CN&7970-8060&4.9&4.9&5.7\\
TiO,FeI&8411-8453&4.9&4.6&4.4\\
TiO,FeI&8453-8490&5.3&5.3&4.4\\
CaII&8490-8508&3.2&3.6&2.9\\
FeI,VO,TiI&8508-8527&3.4&2.9&2.9\\
CaII&8527-8557&6.4&6.6&6.1\\
FeI,VO,TiO&8557-8645&9.2&10.3&9.3\\
CaII&8645-8677&5.4&5.3&4.8\\
FeI,MgI&8677-8768&7.0&8.6&7.2\\
FeI,MgI&8768-8855&3.8&5.2&4.0\\
\noalign{\smallskip}
\hline
\end{tabular}
}
}
\end{table}

\begin{table}
{\scriptsize
\caption{M 81}
\centerline{
\begin{tabular}{|l|c|c|c|c|c|}
\noalign{\smallskip}
\hline
\noalign{\smallskip}
identification& $\lambda$ (\AA)&nucleus&ring 1&ring2\\
\noalign{\smallskip}
\hline
\noalign{\smallskip}
FeI&5058-5156&22.4&20.3&17.4\\
FeI,MgI+MgH&5156-5240&19.6&19.5&17.2\\
FeI&5240-5308&7.7&7.6&6.9\\
FeI&5308-5356&2.9&2.4&2.3\\
FeI&5356-5421&4.0&3.6&3.2\\
FeI,TiO,MgI&5421-5554&9.8&8.2&7.6\\
CaI,FeI,TiO&5554-5630&6.5&5.6&5.1\\
FeI,TiO&5630-5676&3.4&2.9&2.5\\
FeI,NaI&5676-5725&2.6&2.3&2.4\\
FeI,TiO&5725-5825&1.6&2.2&3.2\\
FeI,TiO,CaI&5825-5874&1.2&1.3&1.8\\
NaI&5874-5914&8.1&7.5&6.5\\
FeI,Ti,MnI&5914-6029&6.7&6.3&6.9\\
FeI,CaI&6029-6110&1.6&1.9&2.4\\
FeI&6110-6148&2.0&2.0&2.2\\
TiO,CaI&6148-6208&5.9&6.0&6.2\\
TiO,FeI&6208-6325&4.1&10.9&11.8\\
FeI,CaH,TiO&6325-6374&1.1&3.3&3.2\\
FeI,CaI&6374-6481&1.5&3.3&3.9\\
FeI,TiO,CaI,BaII&6481-6535&-&2.7&2.7\\
$H_{\alpha}$,TiO&6535-6582&-&-&2.1\\
FeI,TiO&6582-6622&-&0.1&1.8\\
TiO&6622-6651&-&1.7&2.2\\
FeI,TiO&6651-6690&3.5&2.5&3.6\\
FeI,TiO,CaI,Si&6690-6761&-&3.8&5.7\\
FeI,TiO,CaH&6761-6795&3.9&2.6&3.0\\
TiO,CaH&6795-6827&3.5&1.8&2.3\\
FeI,SiI&6969-7048&3.9&1.6&2.0\\
TiO&7048-7078&2.6&1.9&2.1\\
TiO,NiI,FeI&7078-7140&9.4&7.8&7.2\\
TiO,VO&7341-7377&1.7&1.6&1.6\\
FeI,VO&7377-7434&2.3&1.9&1.8\\
FeI,VO&7434-7482&1.4&0.9&-\\
FeI,VO&7482-7540&0.9&0.5&-\\
TiO,OI&7737-7823&5.6&6.2&5.9\\
VO,CN,FeI&7823-7888&5.0&4.6&4.6\\
VO,CN,FeI&7888-7970&9.7&8.9&8.4\\
FeI,CN&7970-8060&5.4&5.4&4.4\\
TiO,FeI&8411-8453&4.9&5.0&4.3\\
TiO,FeI&8453-8490&4.9&5.0&4.2\\
CaII&8490-8508&3.4&3.8&3.4\\
FeI,VO,TiI&8508-8527&3.3&2.8&2.4\\
CaII&8527-8557&6.8&6.9&6.0\\
FeI,VO,TiO&8557-8645&8.3&9.2&7.8\\
CaII&8645-8677&5.1&5.9&5.1\\
FeI,MgI&8677-8768&6.8&7.3&6.0\\
FeI,MgI&8768-8855&2.3&3.8&2.7\\
\noalign{\smallskip}
\hline
\end{tabular}
}
}
\end{table}

\begin{table}
{\scriptsize
\caption{NGC 5033}
\centerline{
\begin{tabular}{|l|c|c|c|c|}
\noalign{\smallskip}
\hline
\noalign{\smallskip}
identification& $\lambda$ (\AA)&nucleus&region 1&region 2\\
\noalign{\smallskip}
\hline
\noalign{\smallskip}
FeI&5058-5156&17.3&16.6&13.0\\
FeI,MgI+MgH&5156-5240&15.1&15.0&12.8\\
FeI&5240-5308&5.7&5.1&3.7\\
FeI&5308-5356&1.7&1.4&2.0\\
FeI&5356-5421&2.9&1.9&1.5\\
FeI,TiO,MgI&5421-5554&7.1&5.1&5.2\\
CaI,FeI,TiO&5554-5630&5.9&4.4&4.3\\
FeI,TiO&5630-5676&3.0&2.0&1.9\\
FeI,NaI&5676-5725&2.9&2.1&1.9\\
FeI,TiO&5725-5825&2.4&1.8&2.0\\
FeI,TiO,CaI&5825-5874&-&0.5&1.1\\
NaI&5874-5914&3.7&4.1&4.0\\
FeI,Ti,MnI&5914-6029&5.3&4.8&5.3\\
FeI,CaI&6029-6110&1.7&1.1&1.9\\
FeI&6110-6148&1.4&1.2&1.0\\
TiO,CaI&6148-6208&4.7&3.7&4.9\\
TiO,FeI&6208-6325&6.1&6.6&7.5\\
FeI,CaH,TiO&6325-6374&1.9&1.7&2.9\\
FeI,CaI&6374-6481&3.3&1.9&4.9\\
FeI,TiO,CaI,BaII&6481-6535&-&2.1&3.7\\
$H_{\alpha}$,TiO&6535-6582&-&-&-\\
FeI,TiO&6582-6622&-&-&0.8\\
TiO&6622-6651&-&1.0&1.3\\
FeI,TiO&6651-6690&1.8&1.9&2.2\\
FeI,TiO,CaI,Si&6690-6761&-&3.0&3.3\\
FeI,TiO,CaH&6761-6795&2.3&1.4&1.7\\
TiO,CaH&6795-6827&1.8&1.0&1.9\\
FeI,SiI&6969-7048&1.1&0.1&1.7\\
TiO&7048-7078&0.8&0.9&1.4\\
TiO,NiI,FeI&7078-7140&5.2&5.6&5.9\\
TiO,VO&7341-7377&1.0&0.9&1.6\\
FeI,VO&7377-7434&1.3&1.1&2.0\\
FeI,VO&7434-7482&-&0.0&1.0\\
FeI,VO&7482-7540&-&-&1.0\\
TiO,OI&7737-7823&3.3&4.6&6.6\\
VO,CN,FeI&7823-7888&2.9&3.5&4.6\\
VO,CN,FeI&7888-7970&6.2&5.6&8.1\\
FeI,CN&7970-8060&2.9&2.1&5.2\\
TiO,FeI&8411-8453&3.4&4.7&4.7\\
TiO,FeI&8453-8490&3.8&4.9&5.0\\
CaII&8490-8508&3.1&3.7&3.7\\
FeI,VO,TiI&8508-8527&3.0&2.8&3.0\\
CaII&8527-8557&6.1&6.7&6.9\\
FeI,VO,TiO&8557-8645&7.3&9.8&9.0\\
CaII&8645-8677&4.9&5.9&5.9\\
FeI,MgI&8677-8768&6.7&9.0&10.0\\
FeI,MgI&8768-8855&2.8&5.5&7.0\\
\noalign{\smallskip}
\hline
\end{tabular}
}
}
\end{table}

\begin{table}
{\scriptsize
\caption{NGC 2110}
\centerline{
\begin{tabular}{|l|c|c|c|}
\noalign{\smallskip}
\hline
\noalign{\smallskip}
identification& $\lambda$ (\AA)&nucleus&ring 1\\
\noalign{\smallskip}
\hline
\noalign{\smallskip}
FeI&5058-5156&20.1&22.6\\
FeI,MgI+MgH&5156-5240&-&-\\
FeI&5240-5308&4.9&9.3\\
FeI&5308-5356&1.7&4.2\\
FeI&5356-5421&2.3&4.2\\
FeI,TiO,MgI&5421-5554&6.5&8.0\\
CaI,FeI,TiO&5554-5630&3.6&4.0\\
FeI,TiO&5630-5676&2.7&3.3\\
FeI,NaI&5676-5725&2.0&3.0\\
FeI,TiO&5725-5825&-&0.7\\
FeI,TiO,CaI&5825-5874&0.6&1.3\\
NaI&5874-5914&4.1&3.3\\
FeI,Ti,MnI&5914-6029&6.2&5.7\\
FeI,CaI&6029-6110&0.1&-\\
FeI&6110-6148&1.4&0.7\\
TiO,CaI&6148-6208&4.4&3.1\\
TiO,FeI&6208-6325&-&-\\
FeI,CaH,TiO&6325-6374&-&-\\
FeI,CaI&6374-6481&-&-\\
FeI,TiO,CaI,BaII&6481-6535&-&-\\
$H_{\alpha}$,TiO&6535-6582&-&-\\
FeI,TiO&6582-6622&-&-\\
TiO&6622-6651&-&-\\
FeI,TiO&6651-6690&-&-\\
FeI,TiO,CaI,Si&6690-6761&-&-\\
FeI,TiO,CaH&6761-6795&3.0&2.4\\
TiO,CaH&6795-6827&2.7&1.8\\
FeI,SiI&6969-7048&1.7&1.0\\
TiO&7048-7078&0.9&1.1\\
TiO,NiI,FeI&7078-7140&3.8&4.7\\
TiO,VO&7341-7377&-&1.0\\
FeI,VO&7377-7434&-&0.7\\
FeI,VO&7434-7482&-&0.4\\
FeI,VO&7482-7540&-&-\\
TiO,OI&7737-7823&3.2&3.5\\
VO,CN,FeI&7823-7888&3.2&3.2\\
VO,CN,FeI&7888-7970&5.9&5.1\\
FeI,CN&7970-8060&2.6&1.5\\
TiO,FeI&8411-8453&3.5&4.0\\
TiO,FeI&8453-8490&4.2&4.2\\
CaII&8490-8508&3.3&2.8\\
FeI,VO,TiI&8508-8527&2.3&2.7\\
CaII&8527-8557&5.9&6.1\\
FeI,VO,TiO&8557-8645&2.6&6.4\\
CaII&8645-8677&4.3&4.2\\
FeI,MgI&8677-8768&3.6&5.8\\
FeI,MgI&8768-8855&0.4&1.9\\
\noalign{\smallskip}
\hline
\end{tabular}
}
}
\end{table}

\begin{table}
{\scriptsize
\caption{Mrk 3}
\centerline{
\begin{tabular}{|l|c|c|}
\noalign{\smallskip}
\hline
\noalign{\smallskip}
identification& $\lambda$ (\AA)&ring 1\\
\noalign{\smallskip}
\hline
\noalign{\smallskip}
FeI&5058-5156&16.8\\
FeI,MgI+MgH&5156-5240&14.4\\
FeI&5240-5308&5.6\\
FeI&5308-5356&2.0\\
FeI&5356-5421&4.0\\
FeI,TiO,MgI&5421-5554&10.4\\
CaI,FeI,TiO&5554-5630&6.2\\
FeI,TiO&5630-5676&5.4\\
FeI,NaI&5676-5725&5.2\\
FeI,TiO&5725-5825&7.0\\
FeI,TiO,CaI&5825-5874&2.8\\
NaI&5874-5914&6.4\\
FeI,Ti,MnI&5914-6029&7.3\\
FeI,CaI&6029-6110&1.3\\
FeI&6110-6148&2.0\\
TiO,CaI&6148-6208&6.2\\
TiO,FeI&6208-6325&-\\
FeI,CaH,TiO&6325-6374&-\\
FeI,CaI&6374-6481&-\\
FeI,TiO,CaI,BaII&6481-6535&-\\
$H_{\alpha}$,TiO&6535-6582&-\\
FeI,TiO&6582-6622&-\\
TiO&6622-6651&-\\
FeI,TiO&6651-6690&-\\
FeI,TiO,CaI,Si&6690-6761&-\\
FeI,TiO,CaH&6761-6795&2.0\\
TiO,CaH&6795-6827&1.7\\
FeI,SiI&6969-7048&1.6\\
TiO&7048-7078&1.7\\
TiO,NiI,FeI&7078-7140&8.0\\
TiO,VO&7341-7377&1.5\\
FeI,VO&7377-7434&1.8\\
FeI,VO&7434-7482&0.6\\
FeI,VO&7482-7540&0.4\\
TiO,OI&7737-7823&6.4\\
VO,CN,FeI&7823-7888&3.5\\
VO,CN,FeI&7888-7970&6.9\\
FeI,CN&7970-8060&4.2\\
TiO,FeI&8411-8453&6.0\\
TiO,FeI&8453-8490&5.6\\
CaII&8490-8508&3.1\\
FeI,VO,TiI&8508-8527&3.9\\
CaII&8527-8557&7.1\\
FeI,VO,TiO&8557-8645&10.2\\
CaII&8645-8677&6.0\\
FeI,MgI&8677-8768&8.5\\
FeI,MgI&8768-8855&7.5\\
\noalign{\smallskip}
\hline
\end{tabular}
}
}
\end{table}

\begin{table}
{\scriptsize
\caption{Mrk 620}
\centerline{
\begin{tabular}{|l|c|c|c|}
\noalign{\smallskip}
\hline
\noalign{\smallskip}
identification& $\lambda$ (\AA)&nucleus&ring 1\\
\noalign{\smallskip}
\hline
\noalign{\smallskip}
FeI&5058-5156&12.2&7.1\\
FeI,MgI+MgH&5156-5240&10.6&8.8\\
FeI&5240-5308&4.9&2.9\\
FeI&5308-5356&1.6&1.1\\
FeI&5356-5421&2.3&1.3\\
FeI,TiO,MgI&5421-5554&3.3&3.8\\
CaI,FeI,TiO&5554-5630&1.7&2.9\\
FeI,TiO&5630-5676&0.4&1.0\\
FeI,NaI&5676-5725&0.3&1.7\\
FeI,TiO&5725-5825&-&2.5\\
FeI,TiO,CaI&5825-5874&-&1.1\\
NaI&5874-5914&4.4&5.6\\
FeI,Ti,MnI&5914-6029&1.7&5.3\\
FeI,CaI&6029-6110&-&1.7\\
FeI&6110-6148&1.1&2.3\\
TiO,CaI&6148-6208&3.5&4.5\\
TiO,FeI&6208-6325&2.7&8.0\\
FeI,CaH,TiO&6325-6374&-&1.6\\
FeI,CaI&6374-6481&4.0&1.2\\
FeI,TiO,CaI,BaII&6481-6535&3.6&-\\
$H_{\alpha}$,TiO&6535-6582&-&-\\
FeI,TiO&6582-6622&-&-\\
TiO&6622-6651&1.6&-\\
FeI,TiO&6651-6690&1.6&-\\
FeI,TiO,CaI,Si&6690-6761&-&-\\
FeI,TiO,CaH&6761-6795&1.5&0.9\\
TiO,CaH&6795-6827&1.3&0.8\\
FeI,SiI&6969-7048&0.6&-\\
TiO&7048-7078&0.6&0.9\\
TiO,NiI,FeI&7078-7140&3.2&4.0\\
TiO,VO&7341-7377&1.3&1.9\\
FeI,VO&7377-7434&1.6&2.4\\
FeI,VO&7434-7482&1.2&1.4\\
FeI,VO&7482-7540&0.9&2.1\\
TiO,OI&7737-7823&-&7.5\\
VO,CN,FeI&7823-7888&4.8&5.1\\
VO,CN,FeI&7888-7970&8.1&7.9\\
FeI,CN&7970-8060&5.5&5.1\\
TiO,FeI&8411-8453&4.1&4.1\\
TiO,FeI&8453-8490&4.5&3.9\\
CaII&8490-8508&3.6&3.4\\
FeI,VO,TiI&8508-8527&2.9&2.3\\
CaII&8527-8557&6.3&6.3\\
FeI,VO,TiO&8557-8645&8.4&7.8\\
CaII&8645-8677&5.0&5.4\\
FeI,MgI&8677-8768&7.2&7.2\\
FeI,MgI&8768-8855&3.1&3.1\\
\noalign{\smallskip}
\hline
\end{tabular}
}
}
\end{table}

\begin{table}
{\scriptsize
\caption{NGC 1275}
\centerline{
\begin{tabular}{|l|c|c|c|}
\noalign{\smallskip}
\hline
\noalign{\smallskip}
identification& $\lambda$ (\AA)&region 1a&region 1b\\
\noalign{\smallskip}
\hline
\noalign{\smallskip}
FeI&5058-5156&4.3&5.0\\
FeI,MgI+MgH&5156-5240&5.2&4.9\\
FeI&5240-5308&1.6&2.2\\
FeI&5308-5356&0.4&0.7\\
FeI&5356-5421&1.3&1.9\\
FeI,TiO,MgI&5421-5554&5.7&6.7\\
CaI,FeI,TiO&5554-5630&2.8&4.1\\
FeI,TiO&5630-5676&1.4&2.2\\
FeI,NaI&5676-5725&1.7&2.8\\
FeI,TiO&5725-5825&2.7&4.4\\
FeI,TiO,CaI&5825-5874&1.8&2.0\\
NaI&5874-5914&4.7&3.7\\
FeI,Ti,MnI&5914-6029&7.9&10.2\\
FeI,CaI&6029-6110&4.1&5.8\\
FeI&6110-6148&2.2&3.1\\
TiO,CaI&6148-6208&4.5&6.3\\
TiO,FeI&6208-6325&-&-\\
FeI,CaH,TiO&6325-6374&-&-\\
FeI,CaI&6374-6481&6.7&9.6\\
FeI,TiO,CaI,BaII&6481-6535&-&-\\
$H_{\alpha}$,TiO&6535-6582&-&-\\
FeI,TiO&6582-6622&-&-\\
TiO&6622-6651&-&1.5\\
FeI,TiO&6651-6690&1.3&2.1\\
FeI,TiO,CaI,Si&6690-6761&-&-\\
FeI,TiO,CaH&6761-6795&1.5&1.6\\
TiO,CaH&6795-6827&1.2&1.8\\
FeI,SiI&6969-7048&2.0&3.1\\
TiO&7048-7078&1.2&1.7\\
TiO,NiI,FeI&7078-7140&4.7&6.3\\
TiO,VO&7341-7377&1.1&1.1\\
FeI,VO&7377-7434&1.8&2.3\\
FeI,VO&7434-7482&0.1&0.8\\
FeI,VO&7482-7540&0.2&1.4\\
TiO,OI&7737-7823&3.5&5.1\\
VO,CN,FeI&7823-7888&3.3&4.7\\
VO,CN,FeI&7888-7970&5.3&7.7\\
FeI,CN&7970-8060&3.4&5.1\\
TiO,FeI&8411-8453&5.4&5.0\\
TiO,FeI&8453-8490&5.2&5.1\\
CaII&8490-8508&3.6&3.4\\
FeI,VO,TiI&8508-8527&3.2&2.9\\
CaII&8527-8557&5.5&5.8\\
FeI,VO,TiO&8557-8645&7.8&9.6\\
CaII&8645-8677&5.3&5.2\\
FeI,MgI&8677-8768&7.6&7.6\\
FeI,MgI&8768-8855&2.4&4.3\\
\noalign{\smallskip}
\hline
\end{tabular}
}
}
\end{table}

\begin{table}
{\scriptsize
\caption{NGC 3516}
\centerline{
\begin{tabular}{|l|c|c|c|}
\noalign{\smallskip}
\hline
\noalign{\smallskip}
identification&$\lambda$ (\AA)&r\'egion 1&r\'egion 1b\\
\noalign{\smallskip}
\hline
\noalign{\smallskip}
FeI&5058-5156&12.3&16.0\\
FeI,MgI+MgH&5156-5240&13.6&14.5\\
FeI&5240-5308&4.5&6.7\\
FeI&5308-5356&0.9&2.8\\
FeI&5356-5421&2.9&3.5\\
FeI,TiO,MgI&5421-5554&6.7&8.3\\
CaI,FeI,TiO&5554-5630&5.1&6.3\\
FeI,TiO&5630-5676&3.2&3.4\\
FeI,NaI&5676-5725&2.6&3.3\\
FeI,TiO&5725-5825&1.1&1.7\\
FeI,TiO,CaI&5825-5874&0.8&0.7\\
NaI&5874-5914&3.8&3.8\\
FeI,Ti,MnI&5914-6029&8.0&6.8\\
FeI,CaI&6029-6110&1.6&1.1\\
FeI&6110-6148&1.8&1.6\\
TiO,CaI&6148-6208&4.1&3.0\\
TiO,FeI&6208-6325&7.5&-\\
FeI,CaH,TiO&6325-6374&2.2&1.5\\
FeI,CaI&6374-6481&3.5&2.5\\
FeI,TiO,CaI,BaII&6481-6535&-&-\\
$H_{\alpha}$,TiO&6535-6582&-&-\\
FeI,TiO&6582-6622&-&-\\
TiO&6622-6651&-&-\\
FeI,TiO&6651-6690&-&-\\
FeI,TiO,CaI,Si&6690-6761&-&-\\
FeI,TiO,CaH&6761-6795&2.3&0.7\\
TiO,CaH&6795-6827&2.6&1.0\\
FeI,SiI&6969-7048&3.1&1.3\\
TiO&7048-7078&2.1&1.6\\
TiO,NiI,FeI&7078-7140&8.5&7.1\\
TiO,VO&7341-7377&2.3&2.4\\
FeI,VO&7377-7434&3.0&3.1\\
FeI,VO&7434-7482&1.9&1.9\\
FeI,VO&7482-7540&2.3&2.9\\
TiO,OI&7737-7823&3.2&5.7\\
VO,CN,FeI&7823-7888&2.9&4.1\\
VO,CN,FeI&7888-7970&4.8&6.7\\
FeI,CN&7970-8060&0.8&2.5\\
TiO,FeI&8411-8453&5.1&4.2\\
TiO,FeI&8453-8490&4.6&4.0\\
CaII&8490-8508&3.4&3.3\\
FeI,VO,TiI&8508-8527&3.1&2.4\\
CaII&8527-8557&7.0&6.2\\
FeI,VO,TiO&8557-8645&10.2&7.4\\
CaII&8645-8677&5.3&5.0\\
FeI,MgI&8677-8768&9.7&5.9\\
FeI,MgI&8768-8855&4.6&1.1\\
\noalign{\smallskip}
\hline
\end{tabular}
}
}
\end{table}


\begin{thebibliography}{}
\bibitem{}Arribas S., Mediavilla E., 1994, ApJ 437, 149 
\bibitem{}Balick B., Heckman T.M., 1981, AA 96, 271 
\\bibitem{}Bender R., Burstein D., Faber S.M., 1992, ApJ 399, 462
\bibitem{}Couture J., Hardy E., 1990, AJ 99, 540
\bibitem{}Davidge T.J., Clark C.C., 1994, AJ 107, 946 
\bibitem{}Davidge T.J., 1992, AJ 103, 1512 
\bibitem{}Davies R.L., Sadler E.M., Peletier R.F., 1993, MNRAS 262, 650 
\bibitem{}Delisle, S., Hardy, E., 1992, AJ 103 711 
\bibitem{}van Dokkum P. G., Franx M., 1995, AJ 110, 2027
\bibitem{}Elvis M., Lockman F., Wilkes B.J., 1989, AJ 97, 777 
\bibitem{}Ferruit P., P\'econtal E., 1994, AA 288, 65 
\bibitem{}Filippenko A.V., Sargent W.L.W., 1985, ApJS 57, 503 
\bibitem{}Goudfrooij P., Hansen L., J$\o$rgensen H.E., N$\o$rgaard-Nielsen 
H. U., de Jong T., van den Hoek L.B., 1994, AAS 104, 179
\bibitem{}Grothues H.G., Schmidt-Kaler Th., 1991, AA 242, 357 
\bibitem{}Heckman T.M., Balick B., 1980, AA 83, 100 
\bibitem{}Holtzman J.A., Faber S.M., Shaya E.J. et al., 1992, AJ 103, 691 
\bibitem{}Howarth I.D., 1983, MNRAS 203, 301 
\bibitem{}Hutsemekers D., Surdej J., 1984, AA 133, 209 
\bibitem{}Johansson L., 1987, AA 182, 179 
\bibitem{}Keel W.C., 1989, AJ 98, 195
\bibitem{}Kennicutt R.C. Jr., 1988, ApJ 334, 144
\bibitem{}Koratkar A., Deustua S.E., Heckman T., Filipenko A.V., Ho L.C., 
Rao M., 1995, ApJ 440, 132
\bibitem{}Marconi G., Matteucci F., Tosi M., 1994, MNRAS 270,35 
\bibitem{}Masegosa J., Moles M., Campos-Aguilar A., 1994, ApJ 420, 576 
\bibitem{}Miller J.S., Goodrich R.W., 1990, ApJ 355, 456 
\bibitem{}Minkowski, R., 1968, AJ 73, 836
\bibitem{}M$\o$ller P., Stiavelli M., Zeilinger W.W., 1995, MNRAS 276, 979 
\bibitem{}Oke J.B., Gunn J.E., 1983, ApJ 266, 713 
\bibitem{}Pastoriza M.G., Dottori H.A., Terlevich E., 
Terlevich R., Diaz A.I., 1993, MNRAS 260, 177 
\bibitem{}Pelat D., 1997, MNRAS 284, 365
\bibitem{}Peletier R.F.,  Davies R.L., Illingworth G.D., Davis L.E., 
Cawson M., 1990, AJ 100, 1091
\bibitem{}Pierce A.K., Breckinridge J.B., 1973, KPNO Contribution no.559,
   ``The Kitt Peak table of photographic solar spectrum wavelengths''
\bibitem{}Scoville, N., 1992, Relationships between active galactic nuclei and starbursts galaxies, ASP Conf. Series, vol. 31, 159 
\bibitem{}Serote Roos M., Boisson C., Joly M., Pelat, D., 1997, MNRAS subm. Paper II
\bibitem{}Serote Roos M., Boisson C., Joly M., Ward M.J., 1996, 
MNRAS 278, 897 
\bibitem{}Spinoglio L., Malkan M.A., 1989, ApJ 342, 83
\bibitem{}Stark A.A., Heiles C., Bally J., Linke R., 1984 Bell Lab. (privately
            distributed tape) 
\bibitem{}Stone R.P.S., 1977 ApJ 218, 767 
\bibitem{}Terlevich, E., Diaz, A.I., Terlevich, R., 1990,  MNRAS 242, 271
\bibitem{}Terndrup, D.M., Davies, R.L., Frogel, J.A., DePoy, D.L., Wells, L.A., 1994, ApJ 432, 518
\bibitem{}Tran, H.D., 1995, ApJ 440, 578 
\bibitem{}Turner T.J., Pounds K.A., 1989 MNRAS 240, 858 
\bibitem{}Vreux, J.M., Dennefeld, M., Andrillat, Y., 1983, AAS 54, 437
\bibitem{}Wade, R.A., Horne, K., 1988, ApJ 324, 411
\bibitem{}Walsh J.R., Roy J.R., 1993, MNRAS 262,27 
\bibitem{}Yamada T., 1994 ApJ 423, L30 
\bibitem{}Zhou X., 1991, AA 248, 367 

\end{thebibliography}
\end{document}